\documentclass[prl,showpacs,twocolumn,superscriptaddress]{revtex4}

\usepackage{graphicx}
\usepackage{color}

\begin{document}

\title{The SrTiO$_3$ displacive transition revisited by Coherent X-ray Diffraction}
             
\author{S. Ravy}
\affiliation{Synchrotron SOLEIL, L'Orme des merisiers, Saint-Aubin BP 48, 
91192 Gif-sur-Yvette cedex, France}
\author{D. Le Bolloc'h}
\affiliation{Laboratoire de physique des solides, Univ. Paris-sud, CNRS, UMR 8502,
F-91405 Orsay Cedex, France}
\author{R. Currat}
\affiliation{Institut Laue-Langevin, 6, rue Jules Horowitz, Bo\^ite postale 156, 
38042 Grenoble Cedex 9, France}
\author{A. Fluerasu}
\author{C. Mocuta}
\affiliation{European Synchrotron Radiation Facility, 6 rue Jules Horowitz,
Bo\^ite postale 220, 38043 Grenoble Cedex, France}
\author{B. Dkhil}
\affiliation{Laboratoire Structures, Propri\'et\'es et Mod\'elisation des Solides, 
Ecole Centrale Paris, CNRS, UMR 8580, Grande Voie des Vignes, 
F-92295 Chatenay-Malabry Cedex, France}

\begin{abstract}
We present a Coherent X-ray Diffraction study of the antiferrodistortive displacive transition
of SrTiO$_3$, a prototypical example of a phase transition for which the critical fluctuations exhibit 
two length scales and two time scales.
From the microbeam x-ray coherent diffraction patterns, we show that the broad (short-length scale) and the 
narrow (long-length scale) components can be spatially disentangled, due to 100 $\mu$m-scale spatial 
variations of the latter.
Moreover, both components exhibit a speckle pattern, which is static on a $\sim$10 mn time-scale. 
This gives evidence that the narrow component corresponds to static ordered domains.
We interpret the speckles in the broad component as due to a very slow dynamical process, 
corresponding to the well-known \emph{central} peak seen in inelastic neutron scattering.
\end{abstract}


\maketitle

Although most issues concerning the application of scaling theory to structural phase transitions have been 
settled long ago (see  {\it e.g.} \cite{brucecowley}), two frequently observed scattering features 
remain unaccounted for within standard scaling theory: the "neutron" central peak (CP) and the "x-ray" 
narrow component (NC). 
Remarkably, both features were first evidenced in studies of the critical behavior associated 
with the  $T_c$=100-105 K antiferrodistortive transition in the perovskite SrTiO$_3$.

The first issue concerns the time scale of the critical fluctuations. 
While far above $T_c$, the fluctuations time scale is governed by the inverse soft-phonon 
frequency, a few degrees above $T_c$ a narrow \emph{central} ({\it i.e.} zero-frequency) line 
appears in the inelastic neutron scattering data \cite{riste,shapiro}, whose weight grows 
critically on approaching T$_c$. 
Its frequency width $\Delta \nu$ is too small to be resolved by 
neutron techniques \cite{topler}, but EPR measurements \cite{reiter} have set an upper 
bound of $\Delta \nu<$0.6 MHz. 
There is substantial evidence \cite{hastings,halperin} that the CP phenomenon 
is connected with slowly-relaxing or frozen bulk defects, such as vacancies or interstitials, 
but direct measurements of the relaxation time associated with these defects is still missing.

Beside this second \emph{time-scale}, another unresolved issue concerns the occurrence of a second 
\emph{length-scale} in the critical fluctuations.
As previously mentioned, the structural phase transition in SrTiO$_3$ has been 
the first example \cite{andrews} where a narrow Lorentzian-squared ($\mathcal{L}^2$) 
component in the critical x-ray scattering profiles has been observed close to $T_c$, in addition to the usual 
and broader Lorentzian ($\mathcal{L}$) component (BC). 
It is now well established by high resolution x-ray diffraction techniques that
critical fluctuations close to structural and magnetic phase transitions involve two 
distinct pretransitional scattering components \cite{cowley}, each one 
corresponding to a diverging length scale, as $T$ approaches $T_c$ .

Furthermore, it has been shown at least in Ho \cite{thurston} 
and SrTiO$_3$ \cite{hirota,hunnefeld} that the NC is sample 
dependent, and that its critical behavior is different from that of the BC, 
with larger critical exponents.
The NC component was found to depend on surface preparation \cite{watson,hunnefeld},
and to arise from a near surface \emph{skin} region (typically 10 to 100 $\mu$m-thick). 
This reinforced the conclusion that the NC was somehow connected with surface disorder
and many studies concluded that long-range strains localized near the surface 
were responsible for the occurrence of the NC \cite{hirota,hunnefeld,wang}.

Due to the similarities in the two phenomena, the question arose as to
whether the neutron CP was related to the x-ray NC \cite{gibaud}.
From a neutron reinvestigation of the critical scattering of SrTiO$_3$, it was shown 
that both phonon softening and CP were needed to account for the x-ray BC \cite{shirane}.
From this result and the absence of NC in the neutron 
high resolution study, it was concluded that the 
CP was in fact not correlated with the NC \cite{shirane}.
In both phenomena however, it is striking that the time scale was never experimentally
obtained, first because the NC is absent from the neutron data, and second 
because the CP energy width is not resolved.
This is the reason why we have performed Coherent X-ray Diffraction (CXD) on the critical 
scattering of SrTiO$_3$.

With the advent of third generation synchrotron light sources, it has become possible to 
perform x-ray diffraction with a coherent beam \cite{sutton},
a technique which has since proved to be a powerful tool to probe disordered systems.
Indeed, CXD no longer results in smooth ensemble-averaged diffraction patterns but
in speckled patterns, which are related to the exact distribution of scatterers within the
radiation's coherence volume.
Moreover, the temporal evolution of the speckles allows one to access the dynamics of the
system, especially in the 1-10$^{4}$ s range (see e.g. \cite{ludwig}). 
In this paper, we show how CXD experiments performed on SrTiO$_3$ in the vicinity of its
antiferrodistortive phase transition have succeeded in shedding a new light on the CP and NC issues.

\begin{figure}[ht]
\vskip -0.4cm
\includegraphics[width=9cm]{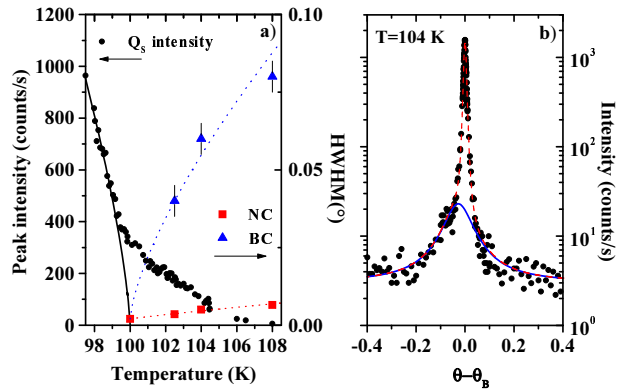}
\vskip -0.4cm
\caption{(Color online) a) Temperature dependence of the peak intensity (circles) of the 
${\bf Q}_s$ superstructure reflection (20$\mu m$$\times$20$\mu m$ entrance slits).
The solid line is a fit to a power law.
Right-hand scale, HWHM of the BC (triangles) and the NC (squares).
The lines are guides for the eye.
b) $\theta$-scan across ${\bf Q}_s$ at 104 K
(200$\mu m$$\times$200$\mu m$ entrance slits).
Intensity scale is logarithmic.
The dashed line is a sum of a $\mathcal{L}$ (solid line) and $\mathcal{L}^{1.7}$ lineshape 
(see text).}
\label{I_T}
\end{figure}

Preliminary experiments have been performed at the ID10a and the ID01 beamlines of the ESRF.
The results presented here were obtained at ID10a, using a Si(111) single crystal monochromator
to get 8 keV x-rays.
A SrTiO$_3$ single-crystal, grown by the top-seeded technique, with a 4mm$\times$4mm 
polished [110] face, was mounted in a top-loading cryostat and aligned with the (311) and 
(1,$\overline{1}$,$\overline{2}$) directions in the horizontal scattering plane.
The geometry of diffraction to reach the ${\bf Q}_s$=($\frac{3}{2}$,$\frac{1}{2}$,$\frac{1}{2}$) 
superstructure reflection 
(Bragg angle $\theta_B$=19.2$^\circ$, Q$_s$=2.68 \AA$^{-1}$) was thus strongly asymmetric 
with the incident angle $\theta_i$=13$^\circ$ and the exit angle $\theta_f$=25.4$^\circ$.
Given the $\mu^{-1}\sim$17~$\mu$m penetration length of 8 keV x-rays, this leads to an effective 
penetration of 3.8~$\mu$m.

The conditions to get a coherent beam were obtained by using 10$\mu$m$\times$10$\mu$m
entrance slits (playing the role of pinhole) 20 cm before the sample.
The beam quality and its intrinsic degree of coherence were tested by using 
2$\mu$m$\times$2$\mu$m entrance slits \cite{lebolloch}, in order to observe their
regular interference fringes in the Fraunhofer regime. 
Guard slits have been placed after the entrance slits to reduce parasitic slit scattering.  
The patterns were recorded on a direct illumination CCD camera (20$\mu$m$\times$20$\mu$m 
pixel size) located 1.8 m after the sample, yielding a resolution of 4.5 10$^{-5}$~\AA$^{-1}$ 
per pixel.

We first checked the phase transition characteristics by measuring the scattering around the 
${\bf Q}_s$ superstructure reflection as a function of temperature, using a point detector and 
20$\mu$m$\times$20$\mu$m detector slits.
The resolution, as determined from the low-temperature Half-Width at Half-Maximum (HWHM) of
the ${\bf Q}_s$ superstructure reflection is 0.002$^\circ$ HWHM (0.9 10$^{-4}$\AA$^{-1}$).
Fig. \ref{I_T}b) shows a $\theta$-scan at 104~K, in which the BC and NC components 
are clearly visible.
In Fig. \ref{I_T}b), the BC has been fitted to a $\mathcal{L}$ lineshape.
The NC, usually fitted to $\mathcal{L}^2$ \cite{hunnefeld,hirota}, 
was found to be better fitted to a $\mathcal{L}^{1.7}$ lineshape.
Finally, Fig. \ref{I_T}b) shows that the BC and NC are shifted by 0.02$^\circ$
($\sim$10$^{-3}$\AA$^{-1}$).
This feature will be discussed later.

The peak intensity at the ${\bf Q}_s$ reciprocal position is displayed in Fig. \ref{I_T}a).
Below $T_c$, a fit to a $(T-T_c)^{2\beta}$ power law assuming $\beta$=0.36,
as expected for a 3D-Heisenberg order parameter,
yields $T_c=$100$\pm$0.2 K, which is consistent with previous studies 
\cite{hunnefeld,hirota}.
Above $T_c$, the behavior of the peak intensity is mainly due to the presence of the 
NC, which is still visible at least 15 K above the phase transition.
The temperature dependence of the BC and NC widths is indicated in Fig. \ref{I_T}a), 
and found to be consistent with previous studies as well \cite{hunnefeld,hirota}.

\begin{figure}[htb]
\vskip -0.4cm
\includegraphics[width=9cm]{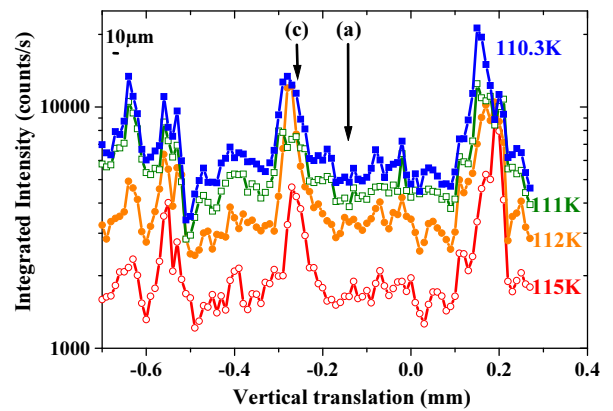}
\vskip -0.4cm
\caption{(Color online) Integrated intensity around the ${\bf Q}_s$ reciprocal position 
as a function of the vertical translation of the sample.
Intensity scale is logarithmic.
The two 2D maps obtained at the positions indicated by arrows are displayed 
in Fig. \ref{speckle}a) and c).}
\label{scan}
\end{figure}

In order to gain some insight into the NC and BC spatial distribution, we have recorded 
x-ray diffraction patterns around ${\bf Q}_s$ while scanning the crystal 
vertical position ($z$) in 10~$\mu$m steps, a value which corresponds to the pinhole size.
The graphs displayed in Fig. \ref{scan} give the patterns integrated intensity 
as a function of $z$ for temperatures between 110 K and 115 K.
These graphs show the presence of 20 to 100 $\mu$m regions in which 
the intensity is several times larger than average.
We checked these results against mosaic effect by performing the same $z$-scans 
at slighly different $\theta$ angles.
Consistently with these results, increasing the beam size (pinhole) to 100 $\mu$m 
made the curves much smoother.

\begin{figure}[htb]
\includegraphics[width=9.5cm]{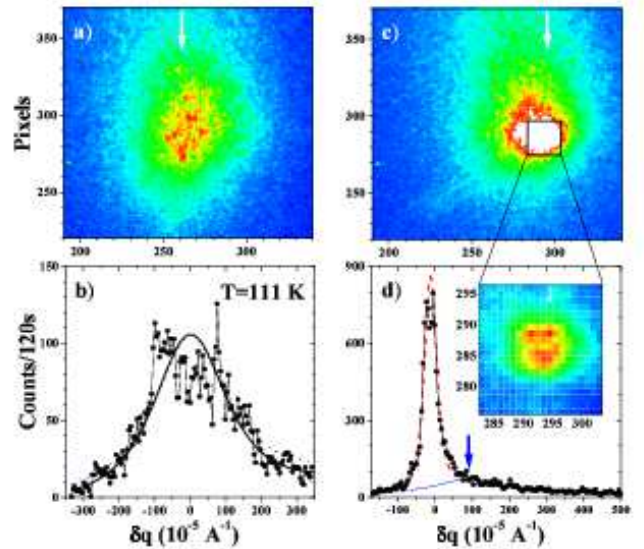}
\vskip -0.4cm
\caption{a) and c) 2D patterns obtained at the  
${\bf Q}_s$ superstructure peak intensity position for two different $z$-positions 
of the beam, corresponding to the arrows in Fig.~\ref{scan}.
The same logarithmic color scale is used for both images.
Saturated values are shown in white.
Figs. b) and d) display the vertical profiles indicated by white arrows in a) and c).
In b) the (dotted) solid lines are fits to a ($\mathcal{L}^2$) $\mathcal{L}$ lineshape.
In d) the dotted line is a sum of a $\mathcal{L}^2$ and a $\mathcal{L}$ 
lineshape (solid line) for the NC and the BC, respectively.
The (blue) arrow points the BC maximum.
The insert in d) is a zoom of c) around the NC, in linear color scale.}
\vskip -0.2cm
\label{speckle}
\end{figure}

The 2D patterns obtained from the CCD camera clearly explain the 
origin of these strong variations: the BC is observed for  
every beam position on the sample (Fig. \ref{speckle}a) and c)),
while the NC dominates the scattering 
in the high intensity regions (Fig. \ref{speckle}c)).
In a few BC dominated regions, $\theta$-scans were performed to check 
for the absence of NC.
The maximum intensity of the BC was found to only vary within a factor 2 
throughout the $z$-scans, as shown in Fig. \ref{scan}.
Finally, the intensity variations shown in Fig. \ref{scan} were also observed 
\emph{below} $T_c$, indicating that the order parameter is more 
developed in regions where the NC is observed above $T_c$.

Consistently with the $\theta$-scan measurements, Fig. \ref{speckle} shows 
that the maxima of the BC and the NC are not at the same position.
The shift is about 10$^{-3}$\AA$^{-1}$ in the vertical direction.
Although it was not discussed there, a similar shift is also apparent in Fig. 3 of 
Ref.\cite{hirota}.
Such an effect was also observed close to the first order antiferrodistortive transition 
in RbCaF$_{3}$ \cite{ryan}.
In both compounds, this shows that the NC and the BC originate from 
regions with different lattice parameters.

This first result gives clear evidence that an inhomogeneous 
distribution of the NC is observed in the sample, and that it is not the case 
for the BC.
Incidentally, this demonstrates that the NC and the BC can be 
disentangled by micro-diffraction, which illustrates the potential value of this technique 
to test phase transitions theories.
 
The second important result of this study is obtained using the 
coherence properties of our X-ray beam: the 2D patterns obtained around 
${\bf Q}_s$ display static speckles on \emph{both} the BC  
and NC, as shown in Figs. \ref{speckle} at T=111 K.
This type of speckle patterns was observed in all regions of the sample we have
studied.
The HWHM of the sharpest speckles is about
$\Delta q\sim$~8$\times$10$^{-5}$\AA$^{-1}$, which gives a direct 
space distance of 2$\pi$/$\Delta q\sim$ 8 $\mu$m. 
This shows that the speckles width is dominated by the pinhole size \cite{note1}.
In Fig. \ref{speckle}, the patterns have been obtained with 2 mn exposure time.
During a longer exposure time of 20 mn, no sizeable evolution of the pattern
was observed.

The implications of this observation are different for each of the two components.
Concerning the NC, the few speckles observed (see insert in Fig. \ref{speckle}d)) 
are due to the diffraction of
a disordered set of quasistatic domains of the low-temperature phase. 
The average size of these domains is given by the HWHM of the NC envelope curve,
which amounts to 2 $\mu$m at T$_c$+11 K.

Different models have been proposed to explain the occurrence of the NC close to 
phase transitions.
Some are based on the presence of quenched defects, while other consider the
phenomenon to be an intrinsic effect \cite{cowley}.
In both cases however, strain fields are thought to be at the origin of the NC.
Considering the absence of NC in large zones of the crystal, it seems very unlikely 
for the NC to result from an intrinsic effect.

In order to explain the universal character of the NC phenomenon, mechanisms 
based on long-range strains have been proposed \cite{altarelli}, following Ref. \cite{weinrib}.
These models rely on the presence of random bonds and/or a small density of impurities, 
which make $T_c$ dependent on position $\bf{r}$ through a local compression or expansion 
of the lattice.
Conditions on the $T_c(\bf{r})$ correlation function make
the transition belong to a "super-universality-class", with critical exponents close
to the experimental values.
Long-range strains are believed to be due to dislocation dipoles, at least in Ho 
(Ref. \cite{thurston}) and Tb (Ref. \cite{gehring}).

Our observations of an inhomogeneous set of quasistatic domains makes the hypothesis
of a nucleation of the low temperature phase around defects, such as dislocations dipoles, 
very plausible.
This is consistent with the NC position being shifted with respect to the BC one \cite{ryan}.
In this respect, it is noteworthy that the 20 to 100 $\mu$m regions of existence of the NC 
are similar in size to the skin thickness reported in Ref. \cite{hunnefeld}.  
However, the definitive proof that the NC occurs around dislocations is still missing.
Such evidence could have been obtained from a concomitant study of the main Bragg 
reflections, but the experimental setup did not enable us to keep the same 
10$\mu m\times 10 \mu m$ zone illuminated during $\theta$ rotations.

The other important observation is the presence 
of static speckles in the BC patterns at all studied temperatures.
In previous studies \cite{shirane}, the BC x-ray scattering has been shown to be 
due to the sum of the soft mode and the CP contribution.
Soft modes frequencies being far beyond the frequencies accessible 
by X-ray intensity fluctuation spectroscopy, they contribute to a 
smooth broad scattering without any speckles \cite{note1}.
Moreover, since close to $T_c$, the CP $q$-dependence has the same HWHM as 
the x-ray BC within 10$\%$ \cite{shirane}, we suggest that the presence of speckles 
in the BC scattering is related to the neutron CP.

As in the case of the NC, two types of theories have attempted to account for the CP,
either through \emph{intrinsic} anharmonic effects or through a 
linear coupling of the soft mode to static (\emph{frozen}) or slowly-relaxing defects 
of appropriate symmetry \cite{halperin}.
The main objection against intrinsic scenarios has always been that a characteristic 
time-scale in the MHz range \cite{reiter} seems difficult to generate, starting from 
phonon frequencies in the THz or GHz range.
The observation of very long-lived speckles on the BC only reinforces this argument.
Concerning length-scales, we find that the BC appears to be homogeneous on a 10
$\mu m$ scale, which implies that the defects involved in the CP are of a different nature
than those which induce the NC.
From neutron data, an estimation of the concentration $c$ of CP-active impurities 
per unit cell \cite{halperin} gives $c$=1.2~10$^{-5}$.
This corresponds to one impurity every 5000~nm$^3$, a volume 
far smaller than our 10$\mu$m beam resolution.
Moreover, the static character of the BC speckles favors the model of \emph{frozen
cell defects} linearly coupled to the order parameter \cite{halperin,note2}. 
However, it does not rule out the additional presence of relaxing cell defects, 
whose contribution to the speckle patterns would be smoothed out by time-averaging.

This study shows that CXD is a powerful tool in order to gain insight into 
unresolved issues near phase transitions, such as the neutron CP and the x-ray NC.
By coupling a microbeam probe with a coherent beam, we demonstrate it is
possible to disentangle the NC and the BC of the critical scattering, and to give evidence
for the quasistatic character of both phenomena.
These results are in favor of surface long-range order defects to explain the NC, and 
of frozen impurities to explain the CP.

We thank A. Moussa\"id, A. Madsen, and G. Baldinozzi for help during the ID10a 
experiments, T. Metzger for help during preliminary experiments at ID01, and
B.~Hehlen and E. Courtens for kindly providing us with the SrTiO$_3$ single crystal.
We are indebted to F. Livet and F. Picca for the coherent diffraction set up and
the data analysis.


\begin{thebibliography}{1}

\bibitem{brucecowley} R.A. Cowley, Adv. Phys. {\bf 29} 1 (1980); 
A.D. Bruce, Adv. Phys. {\bf 29} 111 (1980); A.D. Bruce and R.A. Cowley, 
Adv. Phys. {\bf 29} 219 (1980).
\bibitem{riste} T. Riste, E. Samuelsen, K. Otnes and J. Feder, Solid State Commun. 
{\bf 9}, 1455 (1971).
\bibitem{shapiro} S. Shapiro, J. D. Axe, G. Shirane and T. Riste, Phys. Rev. B {\bf 6}, 
4332 (1972).
\bibitem{topler} J. T\"opler, B. Alefeld and A. Heidemann, J. Phys. C {\bf 10}, 635 (1977).
\bibitem{reiter} G. F. Reiter, W. Berlinger, K. A. M\"uller, P. Heller, 
Phys. Rev. B {\bf 21}, 1 (1980).
\bibitem{hastings} J.B. Hastings, S.M. Shapiro and B.C. Frazer, Phys. Rev. Lett. {\bf 40}, 237 (1978).
\bibitem{halperin} B. I. Halperin and C. M. Varma, Phys. Rev. B {\bf 14}, 4030 (1976).
\bibitem{andrews} S. Andrews, J. Phys. C {\bf 19}, 3721 (1986).
\bibitem{cowley} R.A. Cowley, Physica Scripta {\bf T66}, 24 (1996).
\bibitem{thurston} T. R. Thurston, G. Helgesen, J. P. Hill, D. Gibbs, B. D. Gaulin and P. J. Simpson, 
Phys. Rev. B {\bf 49}, 15 730 (1994).
\bibitem{hirota} K. Hirota, J.P. Hill, S.M. Shapiro, G. Shirane, and Y. Fujii, Phys. Rev. B {\bf 52}, 
13195 (1995).
\bibitem{hunnefeld} H. H\"unnefeld, T. Niem\"oller, J. R. Schneider, U. R\"utt, S. Rodewald, 
J. Fleig and G. Shirane, Phys. Rev. B {\bf 66}, 014113 (2002).
\bibitem{watson} G. M. Watson, B. D. Gaulin, Doon Gibbs, T. R. Thurston, P. J. Simpson, S. M. Shapiro,
G. H. Lander, Hj. Matzke, S. Wang and M. Dudley, Phys. Rev. B {\bf 53}, 686 (1996).
\bibitem{wang} R. Wang, Y. Zhu, and S.M. Shapiro, Phys. Rev. Lett. {\bf 80}, 2370 (1998).
\bibitem{gibaud} A. Gibaud, H. You, S.M. Shapiro and J.Y. Gesland, Phys. Rev. B {\bf 42}, 8255 (1990). 
\bibitem{shirane} G. Shirane, R.A. Cowley, M. Matsuda and S.M. Shapiro, Phys. Rev. B {\bf 48}, 
15595 (1993).
\bibitem{sutton} M. Sutton, S. G. J. Mochrie, T. Greytak, S. E. Nagler, L. E. Berman, 
G. A. Held and G. B. Stephenson, Nature {\bf 352}, 608 (1991).
\bibitem{ludwig} K. Ludwig, F. Livet, F. Bley, J.-P. Simon, R. Caudron, D. Le Bolloc'h and A. Moussaid,
Phys. Rev. B {\bf 72}, 144201 (2005).
\bibitem{lebolloch}  D. Le Bolloc'h, F. Livet, F. Bley, T. Schulli, M. Veron and T.H. Metzger,
J. Synchrotron Rad. (2002). {\bf 9}, 258-265.
\bibitem{ryan} T.W. Ryan, R.J. Nelmes, R.A. Cowley, and A. Gibaud, Phy. Rev. Lett. {\bf 56},
2704, (1986).
\bibitem{note1} The degree of coherence of the speckle pattern shown in Fig. \ref{speckle}a)
is $\beta\sim$3 \%.
This low value is mainly due to the large maximum path length difference 
$\Delta\sim2\mu^{-1}\sin^2 \theta$ ($\sim3.6\mu m$) 
compared to the longitudinal coherence length ($\sim1.4\mu m$) given by the monochromator 
\cite{lebolloch2}.
Note that the smooth soft mode contribution to the BC decreases $\beta$ as well.
\bibitem{lebolloch2} D. Le Bolloc'h, S. Ravy, J. Dumas, J. Marcus, F. Livet, C. Detlefs, F. Yakhou, 
and L. Paolasini, Phys. Rev. Lett. {\bf 95}, 116401 (2005)  
\bibitem{altarelli} M. Altarelli, M.D. N\'u\~nez-Regueiro and M. Papoular, 
Phys. Rev. Lett. {\bf 74}, 3840 (1995); M. Papoular, M.D. N\'u\~nez-Regueiro and M. Altarelli,
Phys. Rev. B {\bf 56}, 166 (1997).
\bibitem{weinrib} A. Weinrib and B.I. Halperin, Phys. Rev. B {\bf 27}, 413 (1983).
\bibitem{gehring} P.M. Gehring, K. Hirota, C.F. Majkrzak, and G. Shirane, Phys. Rev. Lett. {\bf71},
1087 (1993).
\bibitem{note2} Such a random-field behavior is known to add a $\mathcal{L}^2$  
contribution to the scattering above T$_c$.
As shown in Fig. \ref{speckle}b), the BC is too speckled to distinguish between a 
$\mathcal{L}^2$ and a $\mathcal{L}$ line shape.

\end{thebibliography}
\end{document}